\documentclass[aps,prd,twocolumn,superscriptaddress,showpacs,byrevtex,nofootinbib]{revtex4-1}
\pdfoutput=1

\usepackage{graphicx}  % needed for figures
\usepackage{epstopdf}
\usepackage{dcolumn}   % needed for some tables
\usepackage{bm}        % for math
\usepackage{amssymb}   % for math
\usepackage{amsmath}
\usepackage{multirow}
\usepackage{float}
\usepackage{psfrag}

\begin{document}
%\linenumbers

\title{Measurement of the $D_s^+ \to \ell^+\nu_\ell$ branching fractions and the decay constant $f_{D_s^+}$}
%\date{\today}

\author{
{\small
  M.~Ablikim$^{1}$, M.~N.~Achasov$^{9,e}$, X.~C.~Ai$^{1}$,
  O.~Albayrak$^{5}$, M.~Albrecht$^{4}$, D.~J.~Ambrose$^{44}$,
  A.~Amoroso$^{49A,49C}$, F.~F.~An$^{1}$, Q.~An$^{46,a}$,
  J.~Z.~Bai$^{1}$
% ...
  \\
  \vspace{0.2cm}
  (BESIII Collaboration)\\
  \vspace{0.2cm} {\it
    $^{1}$ Institute of High Energy Physics, Beijing 100049, People's Republic of China\\
% ...
$^{4}$ Bochum Ruhr-University, D-44780 Bochum, Germany\\
    $^{5}$ Carnegie Mellon University, Pittsburgh, Pennsylvania 15213, USA\\
% ...
$^{9}$ G.I. Budker Institute of Nuclear Physics SB RAS (BINP), Novosibirsk 630090, Russia\\
% ...
$^{44}$ University of Rochester, Rochester, New York 14627, USA\\
% ...
$^{46}$ University of Science and Technology of China, Hefei 230026, People's Republic of China\\
% ...
$^{49}$ (A)University of Turin, I-10125, Turin, Italy; (B)University of Eastern Piedmont, I-15121, Alessandria, Italy; (C)INFN, I-10125, Turin, Italy\\
% ...
$^{a}$ Also at State Key Laboratory of Particle Detection and Electronics, Beijing 100049, Hefei 230026, People's Republic of China\\
% ...
$^{e}$ Also at the Novosibirsk State University, Novosibirsk, 630090, Russia\\
% ...
}}
\vspace{0.4cm}
}

\begin{abstract}
  Using 482~pb$^{-1}$ of $e^+e^-$ collision data collected at a center-of-mass energy of $\sqrt{s} = 4.009$ GeV with the BESIII
  detector, we measure the
  branching fractions of the decays $D_s^+\to\mu^+\nu_\mu$ and $D_s^+\to\tau^+\nu_\tau$. By constraining the
  ratio of decay rates of $D_s^+$ to $\tau^+\nu_\tau$ and to $\mu^+\nu_\mu$ to the Standard Model prediction, the branching fractions are
  determined to be $\mathcal{B}(D_s^+ \to \mu^+\nu_\mu) = (0.495 \pm 0.067 \pm 0.026)\%$ and
  $\mathcal{B}(D_s^+ \to \tau^+\nu_\tau) = (4.83 \pm 0.65 \pm 0.26)\%$. Using these branching fractions, we obtain a value for
  the decay constant $f_{D_s^+}$ of $(241.0 \pm 16.3 \pm 6.5)~\text{MeV}$, where the first error is statistical and the second
  systematic.
\end{abstract}

\maketitle

\section{\label{sec:intro} Introduction}

\begin{figure}[htbp]
\centering
\includegraphics[width=0.3\textwidth]{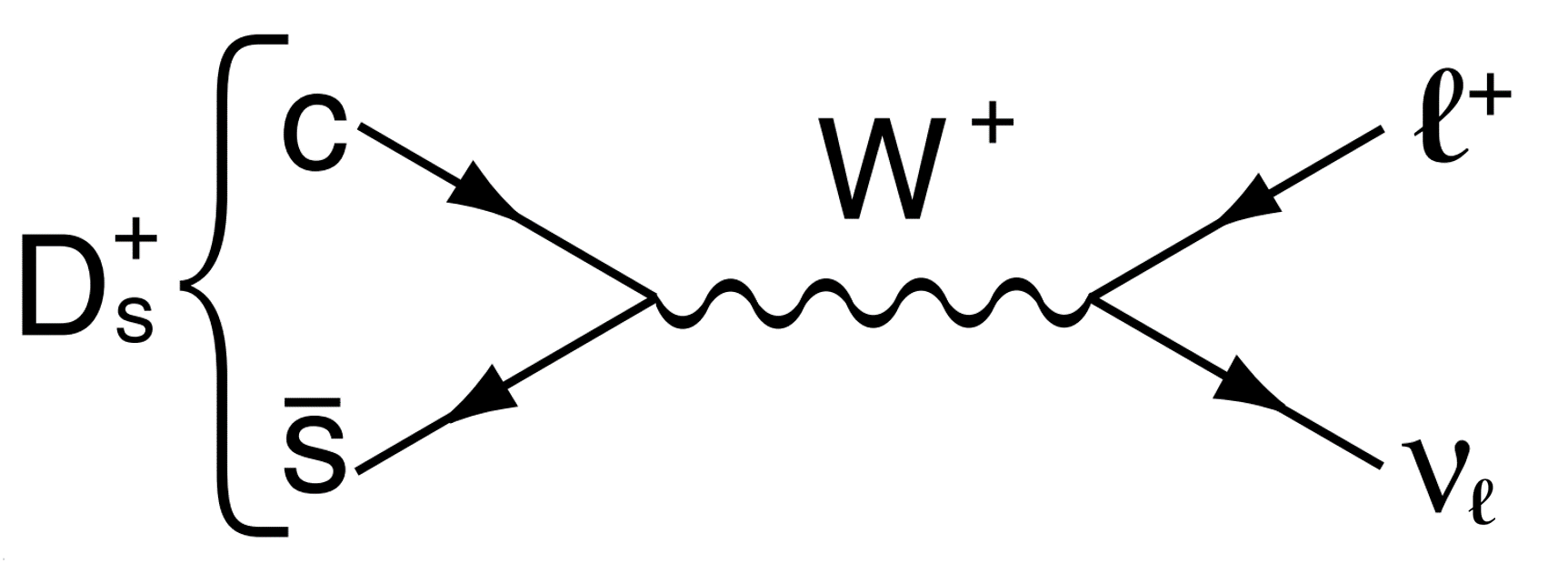}
\caption{Annihilation process for $D_s^+$ leptonic decays in the Standard Model.}
\label{fig:feynman}
\end{figure}

The simplest and cleanest decay modes of the $D_s^+$ meson, both theoretically and experimentally, are the purely leptonic decays. In the
Standard Model (SM), $D_s^+$ leptonic decays proceed via the annihilation of the $c$ and anti-$s$ quarks into a virtual
$W^+$ boson (Fig.~\ref{fig:feynman}). The decay rate is predicted~\cite{Silverman:1988} to be
\begin{equation}
  \Gamma\left(D_{s}^{+} \to \ell^+\nu_\ell\right)=\frac{G_F^2}{8\pi}f_{D_s^+}^2m_\ell^2m_{D_s^+}\left(1-\frac{m_\ell^2}{m_{D_s^+}^2}\right)^2|V_{cs}|^2,
  \label{eq:leptonic width}
\end{equation}
where $m_{D_s^+}$ is the $D_s^+$ mass, $m_\ell$ is the lepton mass, $G_F$ is the Fermi coupling constant, $|V_{cs}|$ is the
Cabibbo-Kobayashi-Maskawa matrix \cite{CKM:1973} element which takes the value equal to $|V_{ud}|$ of $0.97425(22)$
\cite{Hardy:2009}, and $f_{D_s^+}$ is the decay constant that is related to the wave-function overlap of the quark and
antiquark. The $D_s^+$ meson leptonic decay is a process in which a spin-0 meson decays to a left-handed neutrino or a right-handed
antineutrino. According to angular momentum conservation, the lepton $\ell^+$ ($\ell^-$) must be left-handed (right-handed). As a
consequence, the leptonic decay of $D_s^+$ meson is helicity suppressed, which follows from the $m_\ell^2$ dependence of the decay
width. Taking the phase-space factor $(1-m_\ell^2/m_{D_s^+}^2)^2$ into account, the leptonic branching fractions are in the ratio
$e^+\nu_e : \mu^+\nu_\mu : \tau^+\nu_\tau \simeq 2 \times 10^{-5} : 1 : 10$. The decays to $\mu^+\nu_\mu$ and $\tau^+\nu_\tau$ can be
measured experimentally, while $e^+\nu_e$ is beyond the sensitivity of the BESIII experiment.

Recently, the CLEO \cite{Alexander:2009}, \emph{BABAR} \cite{Sanchez:2010}, and Belle \cite{Zupanc:arXiv} collaborations have published
updated
measurements of the branching fractions of $D_s^+$ leptonic decays and the decay constant $f_{D_s^+}$, resulting in the new world
average $f_{D^+_s} = (257.5 \pm 4.6)\;\text{MeV}$ \cite{Olive:2014}. Theoretical predictions of $f_{D_s^+}$
\cite{Carrasco:2015,Bazavov:2014,Yang:2015,Na:2012,Bazavov:2012,Davies:2010} are lower than this
value. The most precise predictions are from lattice QCD; the combined $(2+1)$- and $(2+1+1)$-flavor result is $(249.0 \pm 1.2)\;\text{MeV}$.
There is an approximately two standard-deviation difference between the
experimental average and the lattice QCD calculations. Several models of physics beyond the SM, such as the two-Higgs-doublet model
\cite{Akeroyd:2007} and the $R$-parity-violating model \cite{Akeroyd:2003}, may help to understand this difference. It is important
to further investigate this difference both theoretically and experimentally.

In this paper, we report new measurements of the branching fractions of $D_s^+ \to \mu^+\nu_\mu$ and
$D_s^+ \to \tau^+\nu_\tau$ (where we use the decay $\tau^+ \to \pi^+\bar{\nu_\tau}$) and use them to determine the decay constant
$f_{D_s^+}$. We use $482~\;\text{pb}^{-1}$ \cite{Ablikim:2015} of $e^+e^-$ annihilation data taken at 4.009~GeV with the BESIII detector. At this energy, $D_s$ mesons
are only produced in $D_s^+D_s^-$ pairs and the cross section of $D_s^+D_s^-$ is nearly maximal \cite{Cronin:2009}. As other
processes, such as $D_sD_s^*$ and $D_s^*D_s^*$, are not allowed kinematically, we benefit from the exceptional purity of the
$D_s^+$ sample. Using the technique first introduced by the MARK III Collaboration \cite{Baltrusaitis:1986, Adler:1988}, we
select single-tag events, where either $D_s^+$ or $D_s^-$ is reconstructed, and then reconstruct the leptonic signal on the recoil
side (signal side). In this paper, we choose nine hadronic modes with large branching fractions to reconstruct single-tag events: (a)
$K_S^0K^-$, (b) $K^+K^-\pi^-$, (c) $K^+K^-\pi^-\pi^0$, (d) $K_S^0K^+\pi^-\pi^-$, (e) $\pi^+\pi^-\pi^-$, (f)
$\pi^-\eta~(\eta \to \gamma\gamma)$, (g) $\pi^-\pi^0\eta~(\eta\to\gamma\gamma)$, (h)
$\pi^-\eta'~(\eta' \to \pi^+\pi^-\eta, \eta\to\gamma\gamma)$, and (i) $\pi^-\eta'(\eta' \to \pi^+\pi^-\gamma)$. For convenience,
we denote the single tag as $D^-_s$ and the leptonic decays as $D^+_s$, although charge-conjugate states are also included.

\section{\label{sec:detector} Detector and Monte Carlo}
The BESIII detector \cite{Ablikim:2010} is designed to study hadron spectroscopy and $\tau$-charm physics \cite{Asner:2009}. The cylindrical BESIII is
composed of a helium-gas based drift chamber (MDC), a time-of-flight (TOF) system, a CsI(Tl) electromagnetic calorimeter (EMC),
and a RPC-based muon chamber (MUC), with a superconducting magnet providing a $1.0\;\text{T}$ magnetic field in the central region of the
detector. The MDC covers the polar angle range $|\cos\theta|< 0.93$, with a momentum resolution of 0.5\% for charged particles at 1~GeV/$c$
and 6\%  resolution in the specific energy loss $dE/dx$. The TOF subdetector consists of two parts, the barrel and end cap. The intrinsic time
resolution for the barrel counters is 80~ps, while for the end-cap counters it is 110~ps. The EMC measures energies and positions of electrons
and photons with an energy resolution of 2.5\% (5\%) at an energy of 1~GeV in the barrel (end cap) region. The MUC is designed to have
the ability to identify more than 90\% of muons with momentum over 0.5~GeV, while misidentifying less than 10\% of charged pions
as muons.

We generate two Monte Carlo (MC) simulated samples for background analysis and efficiency measurement. The first sample is a generic MC
sample, which corresponds to an equivalent integrated luminosity of about 20 times the data luminosity and includes open charm processes, continuum production of hadrons, QED processes
and initial-state radiation (ISR) processes. The open-charm processes are simulated at the center-of-mass energy of 4.009~GeV, and
their cross sections are taken from Ref.~\cite{Cronin:2009}. The second sample is an exclusive signal MC sample, in which the
$D_s^-$ meson decays to one of the single-tag modes while the $D_s^+$ meson decays to $\mu^+\nu_\mu$ or
$\tau^+\nu_\tau (\tau^+ \to \pi^+ \bar{\nu_\tau})$. The simulation, including the beam-energy spread, ISR \cite{Kureav:1985} and
final-state radiation \cite{Richter:1993}, is implemented with KKMC \cite{Jadach:2001}. The known decay modes are generated
with EVTGEN \cite{Lange:2001} with branching fractions set to the world average values \cite{Olive:2014}, while the unmeasured
decays are generated with LUNDCHARM \cite{Chen:2000}.

\section{\label{sec:ST selection} Selection of $D_s^-$ single tag}
At $\sqrt{s} = 4.009\;\text{GeV}$,  $D_s$ can only be produced in $D_s^+D_s^-$ pairs. If therefore a $D_s^-$ meson is tagged, the recoil side is guaranteed
to be a $D_s^+$. The $D_s^-$ tag is reconstructed from combinations of charged particles and photons in the event. For charged
particles, the polar angles must satisfy $|\cos\theta|<0.93$, and the points of closest approach to the $e^+e^-$ interaction point
(IP) must be within $\pm10\;\text{cm}$ along the beam direction and within $1\;\text{cm}$ in the plane perpendicular to the beam direction. Charged
pions and kaons must satisfy particle identification (PID) requirements. We calculate the confidence levels for the pion (kaon)
($CL_{\pi(K)}$) hypothesis by combining the ionization energy loss ($dE/dx$) in the MDC and the flight time obtained from the
TOF. The pion (kaon) candidates are required to satisfy $CL_{\pi(K)} > CL_{K(\pi)}$.

For photon candidates, we require that the deposited energy of a neutral shower in the EMC is larger than 25 MeV in the barrel region
($|\cos\theta|<0.8$) or larger than 50 MeV in the end-cap region ($0.86 <|\cos\theta|< 0.92$). To suppress electronic noise and
energy deposits unrelated to the event, the EMC timing of the cluster ($T$) with respect to the event start time is required to satisfy
$0 \leqslant T \leqslant 700$~ns. Photon candidates must be separated by at least 10 degrees from the extrapolated position of any
charged track in the EMC.

The $\pi^0$ and $\eta$ mesons are reconstructed in their $\gamma\gamma$ decay modes. We reject a combination if both photons are detected in
the end-cap of the EMC. The invariant mass of the two photons $M(\gamma\gamma)$ is required to be within $0.115 <M(\gamma\gamma)<
0.150$~GeV/$c^2$ for $\pi^0$ and $0.51 <M(\gamma\gamma)< 0.57$~GeV/$c^2$ for $\eta$, respectively. To improve the resolution, the
$\gamma\gamma$ invariant mass is constrained to the nominal $\pi^0$ or $\eta$ mass \cite{Olive:2014}, and the resultant momenta are
used in the subsequent analysis. The $\eta'$ meson is reconstructed in the $\pi^+\pi^-\eta$ and $\pi^+\pi^-\gamma$ final states. The
invariant masses are required to satisfy $0.943 <M(\pi^+\pi^-\gamma\gamma)< 0.973$~GeV/$c^2$ and $0.932 <M(\pi^+\pi^-\gamma)<
0.980$~GeV/$c^2$ for these two modes, respectively.

Candidates for $K_S^0$ are reconstructed from pairs of oppositely charged tracks without requirements on PID and their distances to the
IP. The secondary vertex is required to be separated from the IP by a decay length of at least twice the vertex resolution. The
invariant mass of the track pair (assuming both tracks are pions) $M(\pi^+\pi^-)$ is required to be within $0.487
< M(\pi^+\pi^-) < 0.511$~GeV/$c^2$.

\begin{figure*}[htbp]
\centering
\includegraphics[width=1.0\textwidth]{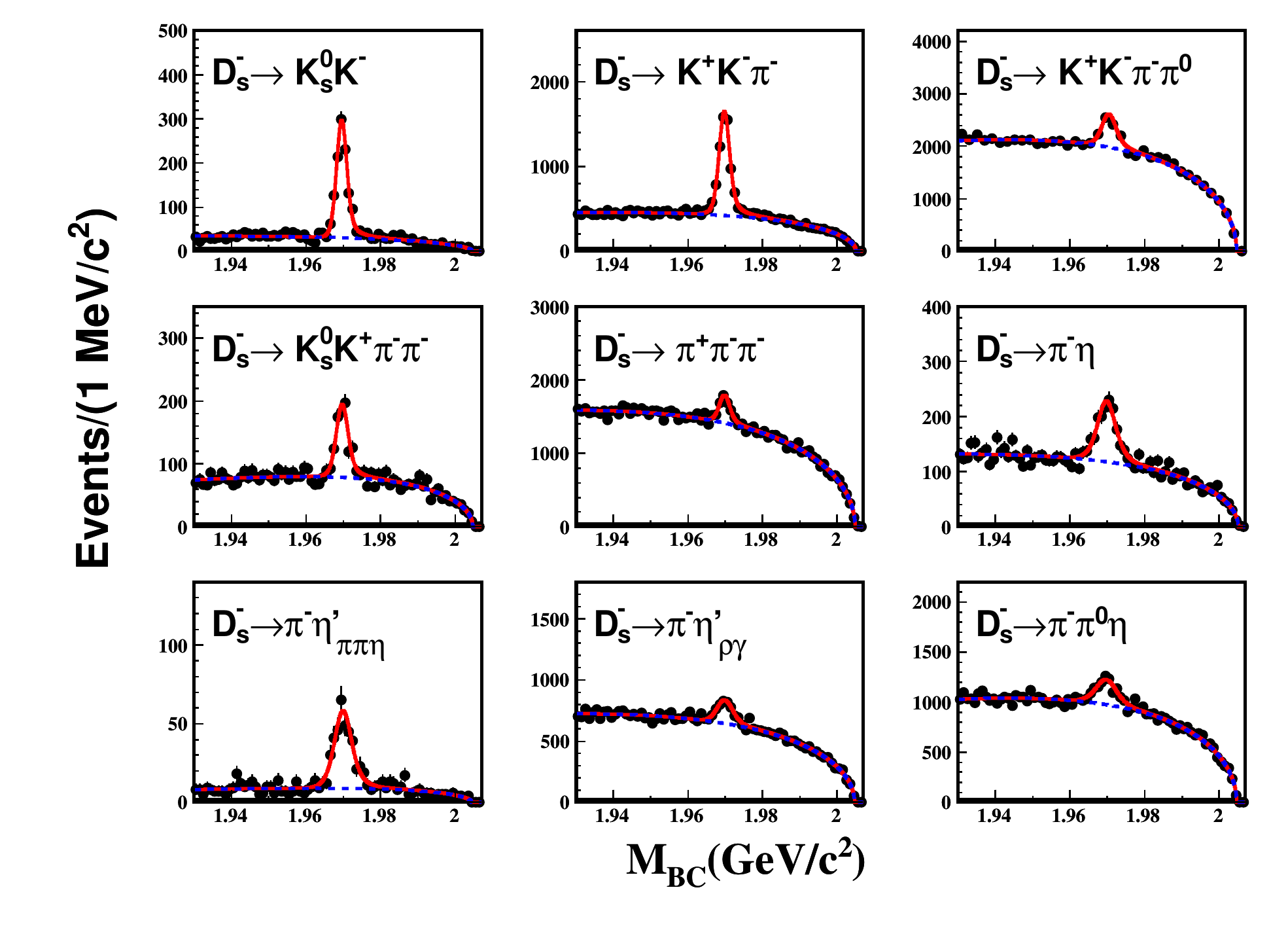}
\caption{Fits to the $M_{\rm BC}$ distributions of $D_s^-$ candidates. The points with error bars are data. The red curves are the
  fit results. The blue dashed curves are the fitted combinatorial backgrounds.}
\label{fig:mbc}
\end{figure*}

Two kinematic variables ($\Delta E$, $M_{\rm{BC}}$) reflecting energy and momentum conservation are used to identify $D_s^-$
candidates. First, we calculate the energy difference
\begin{equation}
  \Delta E = E_{D_s^-} - E_{\rm beam},
\end{equation}
where $E_{D_s^-}$ is the reconstructed energy of a $D_s^-$ meson and $E_{\rm beam}$ is the beam energy. Correctly reconstructed
signal events peak around zero
in the $\Delta E$ distribution. The $\Delta E$ requirements listed in Table \ref{tab:de cuts} cover about 95\% of the signal
events. We keep the combination with the smallest $|\Delta E|$ for each $D_s^-$ tag mode. The second variable
is the beam-energy-constrained mass
\begin{equation}
  M_{\rm BC}=\sqrt{E_{\rm beam}^2/c^4-\overrightarrow{{p}}_{D_s^-}^2/c^2},
\end{equation}
where $\overrightarrow{{p}}_{D_s^-}$ is the total momentum of the particles that form the $D_s^-$ candidate. Figure~\ref{fig:mbc}
shows the $M_{\rm{BC}}$ distributions for data. We determine the single-tag yields by fitting the $M_{\rm{BC}}$
distributions. In the fits, we use the MC-determined signal shapes convolved with a Gaussian function with free mean and resolution
to model the signal and an ARGUS \cite{Albrecht:1989} function for the background. We accept the events satisfying
$1.962 < M_{\rm BC} < 1.982$~GeV/$c^2$ for further analysis. This range contains about 95\% of the signal events. Table
\ref{tab:de cuts} lists the single-tag yields by tag mode, with an overall total of $15127 \pm 321$ $D_s^-$ events.

\begin{table*}[htbp]
  \centering
  \caption{Requirements on $\Delta E$ and $M_{\rm{BC}}$, detection efficiencies and event yields for the different single-tag modes from data (the errors are statistical).}
  \label{tab:de cuts}
  \begin{tabular}{l|c|c|r@{~$\pm$~}l|r@{~$\pm$~}l|r@{~$\pm$~}l|r@{~$\pm$~}l}
    \hline\hline
      Mode & $\Delta E$ (MeV) & $M_{\rm{BC}}$ (MeV) & \multicolumn{2}{c|}{$\epsilon_{\rm{tag}}$ (\%)} & \multicolumn{2}{c|}{$\epsilon_{\rm{tag},\mu\nu}$ (\%)} & \multicolumn{2}{c|}{$\epsilon_{\rm{tag},\tau\nu}$ (\%)} & \multicolumn{2}{c}{$N_{\rm tag}$} \\
    \hline
      $K_S^0K^-$                                & (-27, 21) & \multirow{9}{*}{(1962, 1982)} & 46.76 & 0.34 & 43.97 & 0.22 & 20.14 & 0.18 & 1065 & 39 \\
      $K^+K^-\pi^-$                              & (-32, 23) & & 42.45 & 0.18 & 37.17 & 0.22 & 17.55 & 0.17 & 5172 & 114 \\
      $K^+K^-\pi^-\pi^0$                        & (-41, 22) & & 12.71 & 0.21 & 12.97 & 0.15 & 6.11 & 0.11 & 1900 & 140 \\
      $K_S^0K^+\pi^-\pi^-$                      & (-35, 24) & & 23.37 & 0.36 & 24.21 & 0.19 & 11.50 & 0.14 & 576 & 48 \\
      $\pi^+\pi^-\pi^-$                         & (-36, 23) & & 58.27 & 0.87 & 49.45 & 0.22 & 23.06 & 0.19 & 1606 & 139 \\
      $\pi^-\eta$                               & (-38, 37) & & 46.34 & 0.67 & 42.30 & 0.25 & 19.66 & 0.18 & 814 & 52 \\
      $\pi^-\pi^0\eta$                          & (-35, 27) & & 24.69 & 0.31 & 24.27 & 0.14 & 11.18 & 0.10 & 2172 & 150 \\
      $\pi^-\eta' (\eta' \to \pi^+\pi^-\eta)$   & (-35, 22) & & 27.83 & 0.49 & 24.43 & 0.19 & 11.59 & 0.14 & 440 & 39 \\
      $\pi^-\eta' (\eta' \to \pi^+\pi^-\gamma)$ & (-53, 30) & & 41.83 & 0.86 & 34.54 & 0.21 & 16.28 & 0.17 & 1383 & 143 \\
    \hline\hline
  \end{tabular}
\end{table*}

\section{\label{sec:DT selection} \boldmath Analysis of $D_s^+$ leptonic signal}
\subsection{\label{sec:Signal selection} \boldmath Selection of $D_s^+$ leptonic signal}
In events containing a selected tag candidate, we search for the $D_s^+$ leptonic decays to $\mu^+\nu_\mu$ and
$\tau^+\nu_\tau (\tau^+ \to \pi^+ \bar{\nu_\tau})$ by using the other final-state particles that are not used to
reconstruct the $D_s^-$ tag. We require that there is exactly one good charged track in the signal side, and that the charge of
the track is opposite to the $D_s^-$ tag. The track satisfies the selection criteria (without PID requirements) for charged tracks
given in Sec.~\ref{sec:ST selection}. We also require the energy of the most energetic neutral cluster in the EMC not associated
with the tag $D_s^-$ to be less than 300~MeV to eliminate background events that contain photon(s). If there are multiple
$D_s^+$ candidates in an event, we only keep the one with the $D_s^-$ tag with the smallest $|\Delta E|$ for further analysis.

To characterize the signal events of $D^+_s\to \ell^+\nu_\ell$, the missing mass squared (MM$^2$) is defined as
\begin{equation}
  \text{MM}^{2}=\left(E_{\rm beam}-E_{\mu^+}\right)^2/c^4-\left(-\overrightarrow{p}_{D_s^-}-\overrightarrow{p}_{\mu^+}\right)^2/c^2,
\end{equation}
where $E_{\mu^+}$ and $\overrightarrow{p}_{\mu^+}$ are the energy and momentum of the muon candidate. For
$D_s^+ \to \mu^+ \nu_\mu$ events, the MM$^2$ should peak around zero since there is only one missing neutrino. For
$D^+_s\to \tau^+\nu_\tau (\tau^+\to \pi^+\bar \nu_\tau)$ events, the MM$^2$ (assuming the track is a muon when calculating the
MM$^2$) has a broad structure due to the presence of the two neutrinos. In this study, the signal region considered is
$-0.15 < \rm{MM}^2 < 0.20$~(GeV/$c^2)^2$, where the higher limit is imposed to exclude background events [e.g. $\eta\pi^+$,
$K^0\pi^+$, $\tau^+\nu_\tau (\tau^+ \to \pi^+\pi^0\bar{\nu_\tau}$), etc.] that contribute significantly above 0.20~(GeV/$c^2)^2$.

\subsection{\label{sec:bkg analysis} Background estimation}
Two classes of background events are considered in this analysis. The first one contains $D_s^+$ events in which the single-tag
$D^-_s$ is correctly reconstructed but the signal side is misreconstructed ($\tau^+\to\mu^+\nu_\mu\bar{\nu}_\mu$,
$\tau^+\to\pi^+\pi^0\bar{\nu}_\mu$ and many other $D_s^+$ decays are considered). The second class contains the non-$D_s^+$ background,
which is expected to be a smooth distribution under the $D_s^-$ peak in the $M_{\rm BC}$ spectra. We investigate the real $D_s^+$
background by examining the $D_s^+D_s^-$ events in the generic MC sample with the signal events excluded. After all selection
criteria are imposed, a total of 104 events survive, which is equivalent to $7.0 \pm 0.7$ events for the $482\;\text{pb}^{-1}$ of data. For
the analysis, we fix the shape and size of this background in the MM$^2$ fits. We estimate the contribution of the second class
of background using candidate events in the $M_{\rm BC}$ sideband, which is defined as (1.946, 1.956) GeV/$c^2$ and (1.986, 2.000)
GeV/$c^2$. The background integral in the sideband region is the same as in the signal region.

\subsection{\label{sec:eff} \boldmath $D_s^+$ detection efficiencies}
The overall detection efficiency for $D^+_s\to \ell^+\nu_\ell$ can be expressed as
\begin{equation}
\epsilon = \Sigma_i\left(\frac{N_{\rm tag}^{i}}{N_{\rm tag}} \times \frac{\epsilon_{\rm tag, \rm sig}^{i}}{\epsilon_{\rm tag}^{i}}\right),
\end{equation}
where $N_{\rm tag}^{i}$ is the number of events for single-tag mode $i$, $N_{\rm tag}$ is the number of events for all single-tag
modes, $\epsilon_{\rm tag, \rm sig}^{i}$ is the efficiency of detecting both the single-tag mode $i$ and the leptonic decays, and
$\epsilon_{\rm tag}^{i}$ is the efficiency of detecting the single-tag mode $i$. We determine $\epsilon_{\rm tag, \rm sig}^{i}$ by
analyzing the signal MC sample and $\epsilon_{\rm tag}^{i}$ by analyzing the generic MC sample (Table \ref{tab:de cuts}). The overall signal efficiencies
are measured to be $(91.4 \pm 0.5)\%$ and $(41.0 \pm 0.3)\%$ for $D_s^+ \to \mu^+ \nu_\mu$ and
$D_s^+ \to \tau^+ \nu_\tau (\tau^+ \to \pi^+ \bar{\nu_\tau})$, respectively, where the errors are from MC statistics. It is worth
noting that the large efficiency difference between these two signal channels is mainly caused by the upper limit on MM$^2$.

\subsection{\label{sec:br} Branching fractions}
The branching fraction of the $D_s^+$ leptonic decay is calculated by
\begin{equation}
  \mathcal{B}(D_s^+ \to \ell^+\nu_\ell)=\frac{N_{\rm sig}}{N_{\rm tag}\times \epsilon},
\label{eq:eff}
\end{equation}
where $N_{\rm sig}$ is the number of the signal events that is determined by a fit to the MM$^2$ spectra. In this work, we fit the
MM$^2$ spectra in two different ways, as described in the following sections.

\subsubsection{\label{sec:br1} The SM-constrained fit}
\begin{figure}[htbp]
\centering
\includegraphics[width=0.48\textwidth]{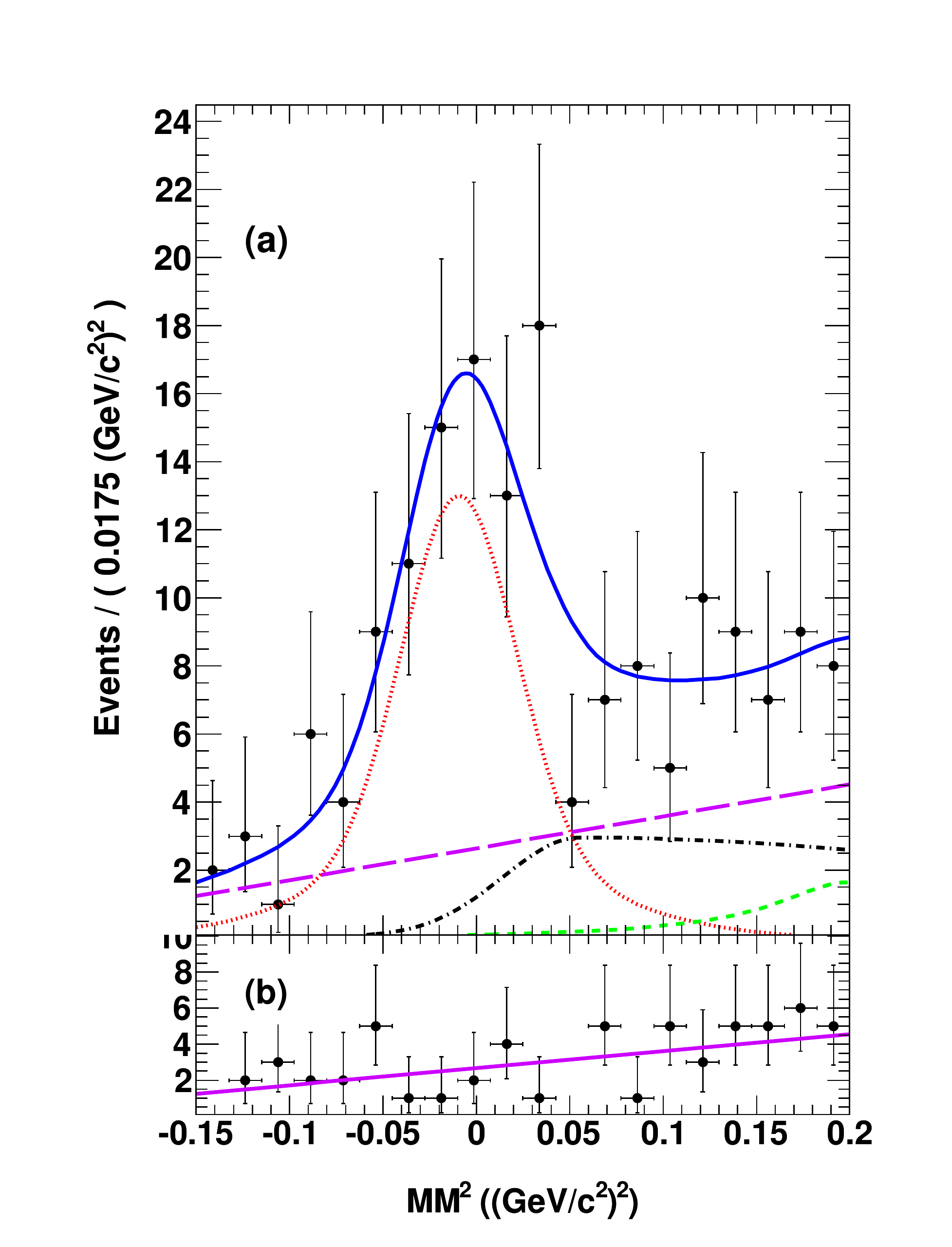}
\caption{Projections of the simultaneous fit to the MM$^2$ distributions of the events in (a) the $D_s^-$ signal region and (b) $M_{\rm{BC}}$ sideband region. Data are shown as the points with error bars. The red dotted curve shows the $\mu^+\nu_\mu$ signal and the black dot-dashed curve shows the $\tau^+\nu_\tau$ signal. The purple long-dashed line shows the non-$D_s^+$ background while the green dashed line shows the real-$D_s^+$ background. The blue curve shows the sum of all these contributions.}
\label{fig:mm2 data1}
\end{figure}

For the finally selected candidates, we fit the MM$^2$ spectra by constraining the ratio of $\mu^+\nu_\mu$ and $\tau^+\nu_\tau$
decay rates to the SM prediction,
\begin{equation}
  R\equiv\frac{\Gamma(D_s^+\to\tau^+\nu_\tau)}{\Gamma(D_s^+\to\mu^+\nu_\mu)}=\frac{m_{\tau^+}^2\left(1-\frac{m_{\tau^+}^2}{m_{D_s^+}^2}\right)^2}{m_{\mu^+}^2\left(1-\frac{m_{\mu^+}^2}{m_{D_s^+}^2}\right)^2}=9.76.
\end{equation}
An unbinned extended maximum likelihood fit to the events in the $M_{\rm{BC}}$ signal region and those in the $M_{\rm BC}$ sideband is
performed simultaneously, as shown in Fig.~\ref{fig:mm2 data1}. In this fit, the ratio of the number of the $\mu^+\nu_\mu$ and
$\tau^+\nu_\tau$ signal events is constrained according to the SM prediction on $R$, the overall signal efficiencies (mentioned in
Sec.~\ref{sec:eff}) and the branching fraction of $\tau^+ \to \pi^+ \bar{\nu_\tau}$. The shapes of the $\mu^+\nu_\mu$ and
$\tau^+\nu_\tau$ signals are determined by the MC shapes convolved with a Gaussian function, the shape and yield of the
real $D^+_s$ background are fixed by the MC estimation, and the non-$D_s^+$ background is modeled by a first-order polynomial function
with parameters and size constrained by the events in the $D^-_s$ sideband in the simultaneous fit. We obtain yields of
$69.3 \pm 9.3$ $D_s^+ \to \mu^+\nu_\mu$ events and $32.5 \pm 4.3$ $D_s^+ \to \tau^+ \nu_\tau (\tau^+ \to \pi^+ \bar{\nu_\tau})$
events, respectively.
Following Dobrescu and Kronfeld's calculation \cite{Dobrescu:2008,Burdman:1995}, we lower the measured
$\mathcal{B}(D_s^+\to\mu^+\nu_\mu)$ by 1\% to account for the contribution of the $\gamma\mu^+\nu_\mu$ final state. The
corrected branching fraction is
\begin{equation}
\mathcal{B}(D_s^+ \to \mu^+\nu_\mu) = (0.495 \pm 0.067)\%,
\label{eq:br munu}
\end{equation}
where the error includes the statistical uncertainties of the single-tag yields and of the signal yield. The corresponding branching fraction of $D_s^+ \to \tau^+ \nu_\tau$ is obtained to be
\begin{equation}
\mathcal{B}(D_s^+ \to \tau^+\nu_\tau) = (4.83 \pm 0.65)\%.
\end{equation}

\begin{figure}[htbp]
\centering
\includegraphics[width=0.4\textwidth]{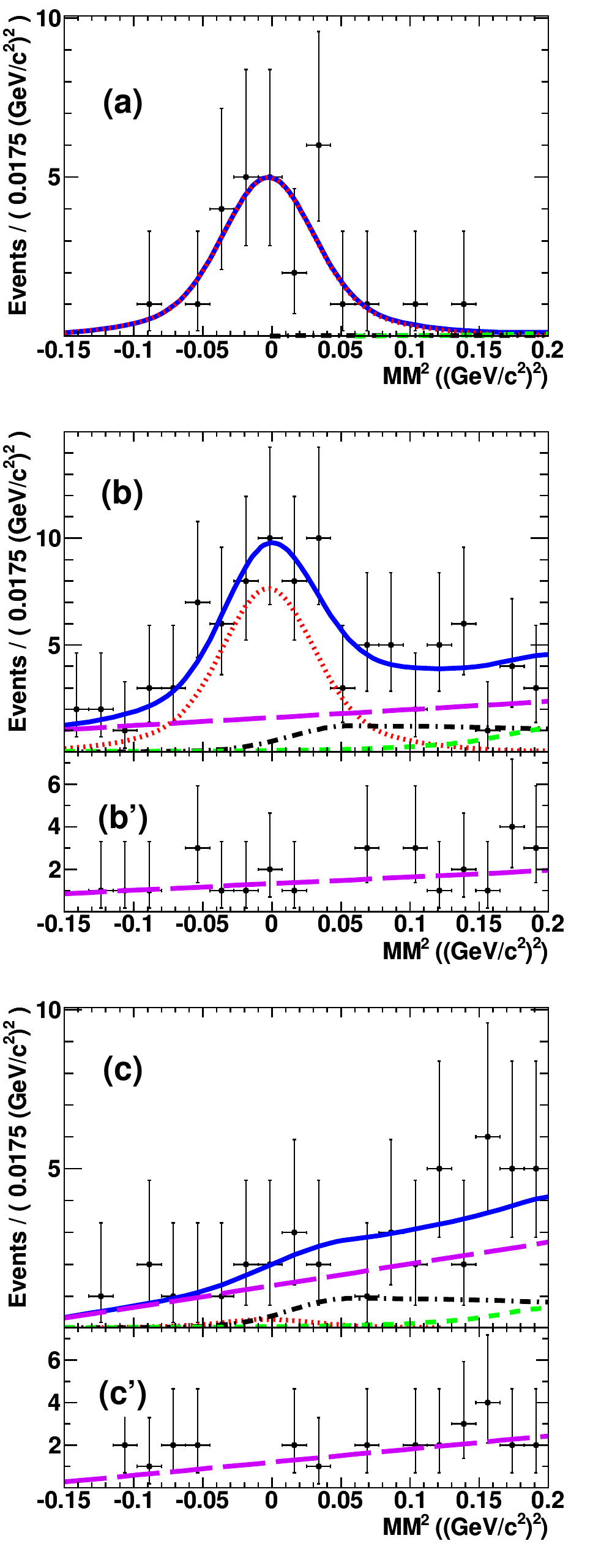}
\caption{Projections of the simultaneous fit to the MM$^2$ distributions of (a) part I, (b) part II and (c) part III data subsamples as defined in Sec.~\ref{sec:br2}. (b') and (c') are the corresponding MM$^2$ distributions from the $M_{\rm BC}$ sideband. Data are shown as the points with error bars. The red dotted curve shows the $\mu^+\nu_\mu$ signal and the black dot-dashed curve shows the $\tau^+\nu_\tau$ signal. The purple long-dashed line shows the non-$D_s^+$ background while the green dashed line shows the real-$D_s^+$ background. The blue curve shows the sum of all these contributions.}
\label{fig:mm2 data2}
\end{figure}

\subsubsection{\label{sec:br2} The non-SM-constrained fit}
Alternatively, we perform a fit to the MM$^2$ spectra leaving the ratio of $\mu^+\nu_\mu$ and $\tau^+\nu_\tau$ events to be free, so that
we can measure the branching fractions of $D_s^+ \to \mu^+ \nu_\mu$ and $D_s^+ \to \tau^+ \nu_\tau$ independently. As shown in
Fig.~\ref{fig:mm2 data1}, it is difficult to distinguish the $\tau^+\nu_\tau$ signal and background in the high MM$^2$ region. We
attempt to improve this situation by taking advantage of the EMC and MUC information.

We use two criteria that help to discriminate muons from pions. In principle, muons can penetrate in the MUC detector much deeper
than hadrons. Therefore, the penetration depth in the MUC can provide strong discrimination power for muons and pions. To select a
muon-enriched sample, we impose the following condition ($\mu\text{-}id$) on the MUC depth $d$: for $p < 1.1~\text{GeV}/c$, we
require $d > (75\,p\,c/\text{GeV} - 40.5)~\text{cm}$, while for $p > 1.1~\text{GeV}/c$, we require $d > 42~\text{cm}$, where $p$
denotes the momentum of the charged track.
This requirement achieves good separation of muons from pions.

The charged tracks deposit energy in the EMC by ionization. For pions, the deposited energies tend to have larger values due to
nuclear interactions in the EMC materials. The condition ($\pi\text{-}id$) to select a pion-enriched sample is
$E_\text{EMC} > 0.3~\text{GeV}$.

We use the above two conditions to separate the $\ell^+\nu_\ell$ candidates into three subsamples. Subsample I contains events
that pass the $\mu\text{-}id$ but fail the $\pi\text{-}id$. Subsample II consists of events that fail both $\mu\text{-}id$ and
$\pi\text{-}id$. Subsample III consists of events that pass the $\pi\text{-}id$. As a result, subsamples I and III are dominated
by muons and pions, respectively, while subsample II has comparable numbers of muons and pions. We measure the relative fractions
of muon (pion) ($\epsilon_{\mu(\pi),\rm{data}}$) in the three subsamples using $e^+e^- \to \mu^+\mu^-$
[$\psi(2S) \to \pi^+\pi^- J/\psi (J/\psi \to \rho\pi)$] events in data. Then we perform a two-dimensional correction (with respect
to momentum and polar angle distributions of the muons or pions in signal MC) to $\epsilon_{\mu(\pi),\rm{data}}$, and obtain the
relative fractions of $\mu^+\nu_\mu$ ($\tau^+\nu_\tau$) ($\epsilon_{\mu\nu(\tau\nu),\rm{data}}$) in the three
subsamples. Table~\ref{tab:sig frac} lists the measured $\mu^+\nu_\mu$ and $\tau^+\nu_\tau$ relative fractions in the three
subsamples in data.

\begin{table}[htbp]
  \centering
  \caption{Relative signal fractions (\%) in the three subsamples (errors are statistical).}
  \label{tab:sig frac}
  \begin{tabular}{lccc}
    \hline\hline
       & I & II & III \\
    \hline
       $\mu^+\nu_\mu$ & $45.6 \pm 0.5$ & $52.9 \pm 0.7$ & $1.9 \pm 0.4$ \\
       $\tau^+\nu_\tau$ & $1.9 \pm 0.1$ & $54.8 \pm 0.6$ & $43.6 \pm 0.6$ \\
    \hline\hline
  \end{tabular}
\end{table}

We perform a simultaneous fit to the MM$^2$ spectra for the events in the three subsamples, constraining the ratio of
$\mu^+\nu_\mu$ to be $45.6:52.9:1.9$ and the ratio of $\tau^+\nu_\tau$ to be $1.9:54.8:43.6$. From the fit, as shown in
Fig. \ref{fig:mm2 data2}, we obtain $72.4 \pm 10.4$ $D_s^+ \to \mu^+\nu_\mu$ events and $22.1 \pm 12.3$
$D_s^+ \to \tau^+\nu_\tau (\tau^+ \to \pi^+\bar{\nu_\tau})$ events.
Applying the correction of 1\%, we find the branching fractions to be
\begin{equation}
  \mathcal{B}(D_s^+ \to \mu^+\nu_\mu)=(0.517 \pm 0.075)\%,
\end{equation}
and
\begin{equation}
  \mathcal{B}(D_s^+ \to \tau^+\nu_\tau)=(3.28 \pm 1.83)\%.
\end{equation}
These results are consistent with those determined from the fit by constraining the $\tau^+\nu_\tau/\mu^+\nu_\mu$ ratio to the SM
prediction. This method can be used to test lepton universality, which demands that the $\tau^+\nu_\tau/\mu^+\nu_\mu$ ratio only
depend on the muon and tau masses.  With the currently available data sample, this test is statistics-limited.

\subsection{\label{sec:syst} Systematic uncertainties}
Table \ref{tab:systematics} summarizes the systematic uncertainties for the branching fraction measurements. The uncertainty due to
the single-tag yield is estimated by varying the fit range and background shape. The uncertainty due to the efficiency of finding
a muon or charged pion is taken to be 1\% per track \cite{Ablikim:2012}. The uncertainty from the efficiency of the extra shower
requirement is studied with the hadronic control samples $\psi(\text{2S})\to\pi^+\pi^- J/\psi (J/\psi \to \mu^+\mu^-)$,
$\psi(\text{2S})\to3(\pi^+\pi^-)$ and $\psi(\text{2S}) \to K^+K^-2(\pi^+\pi^-)$. We fully reconstruct these three samples and
measure the efficiencies for the extra shower requirement for data and MC, respectively. The efficiency difference is taken as the
systematic uncertainty. Uncertainties related to the MM$^2$ fits include the MM$^2$ resolution, MM$^2$ fit range,
background estimation and signal fractions in subsamples. The uncertainty from the MM$^{2}$ resolution is estimated by changing the
resolution of the convolved Gaussian function in signal shape; the uncertainty from the MM$^2$ fit range is estimated by shifting
the range by $\pm 10$~(MeV/$c^2)^2$; the uncertainty due to the background is estimated by varying the number of background events by
$\pm 1\sigma$, assuming that the number of background events follow a Poisson distribution, for the real-$D_s$ background, and
varying the sideband range and background shape for the non-$D_s$ background; the uncertainty from the relative signal fractions in
the subsamples is estimated by varying the fractions by $\pm 1$ statistical error. The systematic error associated with
Dobrescu and Kronfeld's calculation \cite{Dobrescu:2008} of the contribution of the $\gamma\mu^+\nu_\mu$ decay mode could be 1\%
of the lowest-order mechanism for photon momenta below 300 MeV. We take 100\% of this correction value, which is 1\%, as the
systematic error. In addition to these, we have considered uncertainties arising from $\mathcal{B}(\tau^+\to\pi^+\bar{\nu_\tau})$
\cite{Olive:2014} and MC statistics of the detection efficiencies.

\begin{table*}[htbp]
  \centering
  \caption{Systematic uncertainties (\%) for the branching fraction measurements.}
  \label{tab:systematics}
  \begin{tabular}{lcccc}
    \hline\hline
    % after \\: \hline or \cline{col1-col2} \cline{col3-col4} ...
    \multirow{2}{*}{Sources} & \multicolumn{2}{c}{Constrained measurement} & \multicolumn{2}{c}{Unconstrained measurement} \\
    \cline{2-5}
     & $D_s^+ \to \mu^+\nu_\mu$ & $D_s^+ \to \tau^+\nu_\tau$ & $D_s^+ \to \mu^+\nu_\mu$ & $D_s^+ \to \tau^+\nu_\tau$ \\
    \hline
    Number of tags & 1.7 & 1.7 & 1.7 & 1.7 \\
    Track finding & 1.0 & 1.0 & 1.0 & 1.0 \\
    Extra shower cut & 0.5 & 0.5 & 0.5 & 0.5 \\
    MM$^2$ resolution & 2.3 & 2.3 & 2.5 & 5.5 \\
    MM$^2$ fitting range & 1.2 & 1.6 & 1.8 & 0.3 \\
    Background & 4.4 & 4.4 & 2.3 & 9.4 \\
    Relative signal fractions in the three subsamples & - & - & 1.1 & 1.1 \\
    Radiative correction & 1.0 & - & 1.0 & - \\
    $\mathcal{B}(\tau^+ \to \pi^+\bar{\nu_\tau})$ & - & 0.6 & - & 0.6 \\
    MC statistics & 0.5 & 0.6 & 0.5 & 0.6 \\
    \hline
    Sum & 5.6 & 5.7 & 4.6 & 11.2 \\
    \hline\hline
  \end{tabular}
\end{table*}

\subsection{\label{sec:f_Ds} \boldmath Decay constant $f_{D_s^+}$}
The decay constant $f_{D_s^+}$ can be determined using Eq.~(\ref{eq:leptonic width}). By substituting
$\mathcal{B}(D_s^+\to\ell^+\nu_\ell)$=$\tau_{D_s^+}\Gamma(D_s^+\to\ell^+\nu_\ell)$, where $\tau_{D_s^+}$ is the $D_s^+$ lifetime,
we obtain
\begin{equation}
  f_{D_s^+}=\frac{1}{G_Fm_\ell\left(1-\frac{m_\ell^2}{m_{D_s^+}^2}\right)|V_{cs}|}\sqrt{\frac{8\pi\mathcal{B}(D_s^+\to\ell^+\nu_\ell)}{m_{D_s^+}\tau_{D_s^+}}}
\end{equation}

We use the $\mathcal{B}(D^+_s \to \mu^+\nu_\mu)$ result of Eq.~(\ref{eq:br munu}) to calculate the decay constant. Inserting
$G_F$, $m_\mu$, $m_{D^+_s}$, $|V_{cs}|=|V_{ud}|$ $=$ 0.97425(22) \cite{Olive:2014}, and the measured $\mathcal{B}(D^+_s \to \mu^+\nu_\mu)$,
we determine the decay constant to be
\begin{equation}
\label{eq:fDs}
  f_{D_s^+}=(241.0 \pm 16.3 \pm 6.6)~\text{MeV},
\end{equation}
where the first error is statistical and the second systematic. Systematic uncertainties include uncertainties in the measured
branching fractions and the input parameters, and the latter one is dominated by the $D^+_s$ lifetime, which is 0.7\%.

\section{\label{sec:summary} Conclusion}
In this paper, we have measured the branching fractions of $D_s^+ \to \mu^+\nu_\mu$ and $D_s^+ \to \tau^+\nu_\tau$ using 482
pb$^{-1}$ of data taken at 4.009 GeV. Our results within the context of the SM are
\begin{equation}
\mathcal{B}(D_s^+ \to \mu^+\nu_\mu) = (0.495 \pm 0.067 \pm 0.026)\%,
\label{eq:br munu summary}
\end{equation}
and
\begin{equation}
\mathcal{B}(D_s^+ \to \tau^+\nu_\tau) = (4.83 \pm 0.65 \pm 0.26)\%.
\end{equation}

Using these branching fractions, the decay constant $f_{D_s^+}$ is determined as shown in Eq.~(\ref{eq:fDs}).

We have also measured the branching fractions without constraining the $\tau^+\nu_\tau$ and $\mu^+\nu_\mu$ decay rates to the SM
prediction, and the results are
\begin{equation}
\mathcal{B}(D_s^+ \to \mu^+\nu_\mu) = (0.517 \pm 0.075 \pm 0.021)\%,
\end{equation}
and
\begin{equation}
\mathcal{B}(D_s^+ \to \tau^+\nu_\tau) = (3.28 \pm 1.83 \pm 0.37)\%.
\end{equation}

The branching fraction for $D_s^+ \to \mu^+\nu_\mu$ measured in this work is consistent with the  experimental world average
\cite{Olive:2014} within one standard deviation, while the branching fraction for $D_s^+\to \tau^+\nu_\tau$ is about 1.5 standard
deviations lower. The measured decay constant $f_{D_s^+}$ is consistent with the average of the lattice QCD calculations
\cite{Carrasco:2015,Bazavov:2014,Yang:2015,Na:2012,Bazavov:2012,Davies:2010}. With the pure $D_s^+D_s^-$ sample, we provide
an overall competitive result in spite of low statistics. As for the future, BESIII is taking data at $\sqrt{s} = 4.18~\text{GeV}$,
in which $D_sD_s^*$ production is maximal, and we will be able to significantly improve the measurement of the decay constant $f_{D_s^+}$.

\vspace{5 mm}

\centerline{\textbf{\large{Acknowledgements}}}

\vspace{5 mm}

The BESIII Collaboration thanks the staff of BEPCII and the IHEP computing center for their strong support. This work is supported
in part by National Key Basic Research Program of China under Contract No. 2015CB856700, 2009CB825204; National Natural Science Foundation of
China (NSFC) under Contracts Nos. 11125525, 11235011, 11322544, 11335008, 11425524, 11205163, 10935007; the Chinese Academy of Sciences (CAS)
Large-Scale Scientific Facility Program; the CAS Center for Excellence in Particle Physics (CCEPP); the Collaborative Innovation
Center for Particles and Interactions (CICPI); Joint Large-Scale Scientific Facility Funds of the NSFC and CAS under Contracts
Nos. 11179007, U1232201, U1332201; CAS under Contracts Nos. KJCX2-YW-N29, KJCX2-YW-N45; 100 Talents Program of CAS; National 1000
Talents Program of China; INPAC and Shanghai Key Laboratory for Particle Physics and Cosmology; German Research Foundation DFG
under Contracts Nos. Collaborative Research Center CRC 1044, FOR 2359; Istituto Nazionale di Fisica Nucleare, Italy; Koninklijke
Nederlandse Akademie van Wetenschappen (KNAW) under Contract No. 530-4CDP03; Ministry of Development of Turkey under Contract
No. DPT2006K-120470; National Natural Science Foundation of China (NSFC) under Contracts Nos. 11405046, U1332103;
Russian Foundation for Basic Research under Contract No. 14-07-91152; The Swedish Resarch Council;
U. S. Department of Energy under Contracts Nos. DE-FG02-05ER41374, DE-SC-0010504, DE-SC0012069, DESC0010118; U.S. National Science
Foundation; University of Groningen (RuG) and the Helmholtzzentrum fuer Schwerionenforschung GmbH (GSI), Darmstadt; WCU Program of
National Research Foundation of Korea under Contract No. R32-2008-000-10155-0.

%%%%%%%%


\begin{thebibliography}{99}
%
\bibitem{Silverman:1988}
  D. Silverman and H. Yao,
  Phys.\ Rev.\ D {\bf 38}, 214 (1988).

\bibitem{CKM:1973}
  M. Kobayashi and T. Maskawa, Prog. Theor. Phys. {\bf 49}, 652 (1973)

\bibitem{Hardy:2009}
  J.C. Hardy and I.S. Towner,
  Phys. Rev. C {\bf 79}, 055502 (2009).

%\bibitem{Bai:1995}
%  J.Z.~Bai {\it et al.}  [BES Collaboration],
%  Phys.\ Rev.\ Lett.\  {\bf 74}, 4599 (1995).

\bibitem{Alexander:2009}
  J.P.~Alexander {\it et al.}  [CLEO Collaboration],
  Phys.\ Rev.\ D {\bf 79}, 052001 (2009).

\bibitem{Sanchez:2010}
  P.~del Amo Sanchez {\it et al.}  [\emph{BABAR} Collaboration],
  Phys.\ Rev.\ D {\bf 82}, 091103(R) (2010).

\bibitem{Zupanc:arXiv}
  A.~Zupanc {\it et al.}  [Belle Collaboration],
  JHEP {\bf 09}, 139 (2013).

\bibitem{Olive:2014}
  K.A. Olive {\it et al.} [Particle Data Group],
  Chin.\ Phys.\ C {\bf 38}, 090001 (2014).

\bibitem{Carrasco:2015}
  N. Carrasco, P. Dimopoulos, R. Frezzotti, P. Lami, V. Lubicz, F. Nazzaro, E. Picca, L. Riggio, G.C. Rossi, F. Sanfilippo, S. Simula, C. Tarantino,
  Phys. Rev. D {\bf 91}, 054507 (2015).

\bibitem{Bazavov:2014}
  A. Bazavov {\it et al.},
  Phys. Rev. D {\bf 90}, 074509 (2014).

\bibitem{Yang:2015}
  Y.B. Yang, Y. Chen, A. Alexandru, S.J. Dong, T. Draper, M. Gong, F.X. Lee, A. Li, K.F. Liu, Z. Liu, M. Lujan,
  Phys. Rev. D {\bf 92}, 034517 (2015).

\bibitem{Na:2012}
  H. Na, C.T.H. Davies, E. Follana, G.P. Lepage and J. Shigemitsu,
  Phys.\ Rev.\ D {\bf 86}, 054510 (2012).

\bibitem{Bazavov:2012}
  A. Bazavov {\it et al.},
  Phys.\ Rev.\ D {\bf 85}, 114506 (2012).

\bibitem{Davies:2010}
  C.T.H. Davies, C. McNeile, E. Follana, G.P. Lepage, H. Na and J. Shigemitsu,
  Phys.\ Rev.\ D {\bf 82}, 114504 (2010).

%\bibitem{Na:2012}
%  H. Na, C.T.H. Davies, E. Follana, G.P. Lepage and J. Shigemitsu,
%  Phys.\ Rev.\ D {\bf 86}, 054510 (2012).
%
%\bibitem{Bazavov:arXiv}
%  A. Bazavov, C. Bernard, C.M. Bouchard, C. DeTar, M. DiPierro, A.X. El-Khadra, R.T. Evans, E.D. Freeland, E. Gamiz, S. Gottlieb, U.M. Heller, J.E. Hetrick, R. Jain, A.S. Kronfeld, J. Laiho, L. Levkova, P.B. Mackenzie, E.T. Neil, M.B. Oktay, J.N. Simone, R. Sugar, D. Toussaint, and R.S. VandeWater,
%  Phys.\ Rev.\ D {\bf 85}, 114506 (2012).
%
%\bibitem{Blossier:2009}
%  B. Blossier {\it el al.},
%  JHEP {\bf 0907}, 043 (2009).
%
%\bibitem{Bordes:2005}
%  J. Bordes, J. Penarrocha, and K. Schilcher,
%  JHEP {\bf 0511}, 014 (2005).
%
%\bibitem{Lucha:2011}
%  W. Lucha, D. Melikhov, and S. Simula,
%  Phys. Lett. B {\bf701}, 82 (2011).
%
%\bibitem{Narison:2013}
%  S. Narison,
%  Phys. Lett. B {\bf718}, 1321 (2013).
%
%\bibitem{Wang:arXiv}
%  Z.-G. Wang,
%  JHEP {\bf 1310}, 208 (2013).
%
%\bibitem{Gelhausen:2013}
%  P. Gelhausen, A. Khodjamirian, A.A. Pivovarov and D. Rosenthal,
%  Phys. Rev. D {\bf 88}, 014015 (2013).

\bibitem{Akeroyd:2007}
  A.G. Akeroyd and C.H. Chen,
  Phys.\ Rev.\ D {\bf 75}, 075004 (2007).

\bibitem{Akeroyd:2003}
  A.G. Akeroyd and S. Recksiegel,
  Phys.\ Lett.\  B {\bf 554}, 38 (2003).

\bibitem{Ablikim:2015}
  M. Ablikim {\it et al.}  [BESIII Collaboration],
  Chin. Phys. C {\bf 39(9)}, 093001 (2015).

\bibitem{Cronin:2009}
  D. Cronin-Hennessy {\it et al.}  [CLEO Collaboration],
  Phys.\ Rev.\ D {\bf 80}, 072001 (2009).

\bibitem{Baltrusaitis:1986}
  R.M. Baltrusaitis {\it et al.} [MARK-III Collaboration],
  Phys. Rev. Lett. {\bf 56}, 2140 (1986).

\bibitem{Adler:1988}
  J. Adler {\it et al.} [MARK-III Collaboration],
  Phys. Rev. Lett. {\bf 60}, 89 (1988).

\bibitem{Ablikim:2010}
  M. Ablikim {\it et al.} [BESIII Collaboration],
  Nucl. Instrum. Meth. A  {\bf 614}, 345 (2010).

\bibitem{Asner:2009}
  D.M. Asner {\it et al.},
  Int. J. Mod. Phys. A {\bf 24}, 499 (2009).

 \bibitem{Kureav:1985}
   E.A. Kureav and V.S. Fadin,
   Sov. J. Nucl. Phys. {\bf 41}, 466 (1985).

\bibitem{Richter:1993}
  E. Richter-Was, Phys. Lett. B {\bf 303}, 163 (1993).

\bibitem{Jadach:2001}
  S. Jadach, B.F.L. Ward and Z. Was,
  Phys.\ Rev.\ D {\bf 63}, 113009 (2001).

\bibitem{Lange:2001}
  D.J. Lange, Nucl. Instrum. Meth. A {\bf 462}, 152 (2001);
  R.G. Ping, Chin. Phys. C {\bf 320}, 599 (2008).

\bibitem{Chen:2000}
  J.C. Chen, G.S. Huang, X.R. Qi, D.H. Zhang and Y.S. Zhu,
  Phys. Rev. D {\bf 62}, 034003 (2000).

\bibitem{Albrecht:1989}
  H. Albrecht {\it et al.}  [ARGUS Collaboration],
  Phys.\ Lett.\ B {\bf 229}, 304 (1989).

\bibitem{Dobrescu:2008}
  B.A. Dobrescu and A.S. Kronfeld,
  Phys.\ Rev.\ Lett.\  {\bf 100}, 241802 (2008).

\bibitem{Burdman:1995}
  G. Burdman, T. Goldman, and D. Wyler,
  Phys.\ Rev.\ D {\bf 51}, 111 (1995).

\bibitem{Ablikim:2012}
  M. Ablikim {\it et al.}  [BESIII Collaboration],
  Phys.\ Rev.\ D {\bf 86}, 071101 (2012).

\end{thebibliography}
\end{document}